\documentclass[10pt,a4paper,twocolumn]{article}
\usepackage[top=30mm, bottom=30mm, left=15mm, right=15mm]{geometry}
\usepackage{amssymb,amsmath,latexsym}
\usepackage[font={small,it}]{caption}
\usepackage[T1]{fontenc}
\usepackage[utf8]{inputenc}
\usepackage{lmodern}
\usepackage{amsmath}
\usepackage{amssymb}
\usepackage{graphicx}
\usepackage{sectsty}
\usepackage{mathrsfs}
\usepackage{multirow}
\usepackage{tabularx}
\usepackage{array}
\sectionfont{\fontsize{12}{15}\selectfont}
\usepackage[colorlinks=true,
            linkcolor=red,
            urlcolor=blue,
            citecolor=cyan]{hyperref}
\title{Generalized Oppenheimer-Snyder Gravitational Collapse into  Regular Black holes}
\author{F. Shojai$^{1,2}$, A. Sadeghi$^1$ and R. Hassannejad$^1$\\$^1$Department of Physics, University of Tehran, P.O. Box 14395-547\\Tehran, Iran.\\$^2$School of Physics, Institute for Research in Fundamental Sciences (IPM),\\P. O. Box 19395-5531, Tehran, Iran.\\}
\begin{document}
\maketitle
%------------------------------------------------------------------------------------
\begin {abstract}
We shall study the formation of a particular class of regular black holes from the gravitational collapse of a massive star. The inside geometry is described by spatially flat Friedmann-Robertson-Walker metric and the stellar matter is distributed uniformly without any pre-assumption about its equation of state. Our model is a generalization of Oppenheimer-Snyder collapse for regular black holes. We have obtained the density and pressure of star by applying the condition of smooth joining of metrics at the freely falling surface of star. Specifying the regular black holes to Hayward and Bardeen cases, we see that the stellar matter is described by a polytropic equation of state and moreover, for the radius smaller than a certain value, the strong energy condition becomes invalid. Then for both black holes, the interior apparent and event  horizons and also the stellar surface are obtained as functions of the proper time of star. At the end,  we have constructed a new two parametric family of regular black holes jointed smoothly to the flat  Friedmann-Robertson-Walker interior metric of a polytropic star with an arbitrary index.

\end{abstract}
\section{Introduction}
Since the formulation of general relativity (GR) by Albert Einstein, there have been many theoretical and experimental efforts to provide a deep understanding of this theory, its solutions, and physical predictions. One of the most remarkable predictions of GR is the existence of black holes (BHs). These are mysterious objects containing trapped surfaces that are characterized by the negative values of null expansion parameters, and a singularity signaling the breakdown of classical GR \cite{1-4-60Mal}. 
From an experimental point of view, the first images of a BH’s shadow are provided by the Event Horizon Telescope \cite{Tel} and the first observation of the merger of a binary BH is confirmed by the LIGO and Virgo collaborations \cite{Li}. BHs are described only by their mass, angular momentum, and the electric charge according to the no-hair theorem \cite{no}.

The BH singularity is connected to the infinite growth of curvature invariants and it is believed that it is not appear in a full theory of quantum gravity \cite{34-35Mal}. According to the weak cosmic censorship conjecture \cite{kh4-5}, if the stellar matter satisfies the null energy condition, the BH singularity remains hidden behind the event horizon that specifies the causal structure of spacetime. In this way, the singularity is inaccessible to the asymptotic observer. In GR, BHs describe the final state of the classical gravitational collapse of massive stars. However, it is believed that the quantum effects can generate repulsive pressure counteracting the gravitational attraction, such that the singularity is avoided at the end of collapse \cite{34-35Mal}.

To get rid of singularities inside the BHs, one can modify the Einstein-Hilbert action by additional terms arising from vacuum polarization and particle creation \cite{Mukh} or assume that the curvature is restricted by some fundamental value \cite{23-25Fro13,Fro16}. The first attempts in this direction were made by Einstein and Rosen who formulated the wormhole geometry \cite{2RodJcap}. Then, Sakharov proposed $p=-\rho$ as the equation of state for high-density region of a star \cite{1DyCqg05} and Gliner suggested that this equation of state could be the final state of  gravitational collapse \cite{2DyCqg05}. Considering the action of general relativity coupled to the nonlinear electrodynamics, Bardeen \cite{3RodJcap-9-21Kho} suggested the first regular BH and then the regular magnetic and electric BHs are studied in \cite{4-8DyCqg04}.  One of the most popular models of non-singular BHs was proposed by Hayward \cite{54Fro16}. It is a regular modified Schwarzschild BH built by containing a fundamental length scale. At small distances, it reduces to de Sitter metric while it is asymptotically Schwarzschild spacetime. Some generalizations of Hayward metric are proposed in \cite{Fro16}. Other regular BH solutions can be built by gluing an inner regular metric to an exterior BH solution through a smooth junction \cite{33-36Zan} or a thin shell \cite{37-40Zan-Zan}. Up to now, many solutions of regular BHs are proposed in GR \cite{Beato99} or in modified gravity theories \cite{29-31RodJcap}. For a general review of different models of non-singular BHs, one can see \cite{40FroPrd16}.

The regular BH solutions using a de Sitter-like core are proposed by Dymnikova et al \cite{9-16ZanIjmd}.
They have shown that if the energy-momentum tensor of a spherically symmetric static charged BH has the symmetry $T^{0}_{0} = - T ^{1}_{1}$ and the BH metric asymptotically behaves as Reissner--Nordstr\"om metric, then the BH has de Sitter behavior at $r\rightarrow 0$ \cite{9-16ZanIjmd}.
%In general, these BHs have an outer event horizon and an inner Cauchy horizon similar to Reissner--Nordstr\"om BH but with the difference that the %surface $r = 0$ is a time--like regular surface. Since the Cauchy horizon is classically unstable to external perturbations \cite{17Bonano},

%There are some arguments, such as the effect of mass inflation, concerning the observational viability of regular BHs. However, %taking into account the Hawking radiation, recently the authors of \cite{Bonano} have shown that the evaporation process ends with %a regular remnant of Planckian size. 
Another feature of regular BHs is that the matter stress-energy tensor violates the strong energy condition (SEC) somewhere inside the horizon which implies repulsive effects and thus avoidance of singularity \cite{61Mal-11Bal}. These BHs are the static non-singular and non-vacuum solutions of Einstein's equations which effectively present the removal of singularity by quantum gravity effects.  

Now, a question may be raised here. 
%Is it possible to get a regular BH from the gravitational collapse of a massive star? 
Do the regular BH solutions are physically relevant, in the sense that they can be produced by the gravitational collapse of a massive star?
Implying some modifications to the OS collapsing model, one can replace the singularity by a bounce \cite{Bambi}.  Also, using some examples, the authors of \cite{dd1z1} have shown that the regular BHs are not the final products of the quantum corrected gravitational collapse. Therefore,  the only possible way to produce a regular BH is to note that the final product depends on the equation of state of the stellar matter. As a first effort, the authors of \cite{ddMb} have constructed a model using an explicit equation of state with some desired properties. It depends on the radial coordinate and the nature of the stellar matter smoothly changes from a stiff fluid to a cosmological constant acting against the formation of singularity. This produces some de Sitter--like region at the center. Therefore, the general question is, what kind of stellar matter is needed? To answer this question, let us recall that the simplest model of gravitational collapse of dust was developed by Oppenheimer, Snyder (OS) and Datt \cite{6Mal} in which there is a smooth junction through a boundary surface between a collapsing Friedmann-Robertson-Walker (FRW) core and an outer Schwarzschild spacetime. To avoid the singularity, it is required that the gravitational collapse stops at a finite radius because of some kind of repulsive effects.
% Classically, if this radius is larger than the Schwarzschild radius, thus a BH doesn't form and instead, a compact object is produced \cite{25Mal}. %Otherwise, the star may undergo a bounce, i.e., it contracts to a minimum radius and
%then expands. In this case, the repulsive effects are confined to a very small interior region of the star and a BH may appear for an external %observer \cite{34-40Mal}. For the bouncing scenario and the formation of exotic compact remnants from the collapsing process, one can 
%see \cite{41Mal,42Mal}. There are many proposals for such remnants and compact objects formed by gravitational collapse. These can be regular BHs, %gravastars (gravitational vacuum stars) \cite{25Mal,42Mal}, black stars \cite {44Mal}, quark stars \cite{45-46Mal}, boson stars \cite{47-48Mal} and %Planck stars \cite{24-26Kho}. Some of these objects could be distinguished from BHs by imaging of their shadow \cite{171Mal}. 

In this paper, we want to describe the classical gravitational collapse to a typical regular BH metric in which there is an arbitrary function that specifies the short distance behavior of geometry.
We assume that at the star surface, the particles follow the radial geodesics and moreover, the interior metric is smoothly joined to the exterior regular BH metric. The initial velocity of particles is assumed to be zero at spatial infinity. This substantially simplifies our calculations and means that the interior FRW geometry of collapsing star is spatially flat. 
We consider a spherically symmetric star such as the standard OS model. This means that we have ignored the gradients of density and pressure components. Also, we assume that there is no interaction between the background stress-energy (the source of regular BH) and the stellar matter except for the gravitational one. \\
Considering a collapsing ball of matter without any special assumption on its equation of state, here we find the star density and pressure as some functions of the proper time of collapse. We also see that the pure radial pressure is zero at the surface of the star which shows the consistency of our assumption. Then for two special regular spacetimes, Hayward and  Bardeen BHs, we have shown that the stellar equation of state is of a polytropic type, resulted by smooth joining the interior and exterior spacetimes. However, as expected, the SEC is not satisfied for the stellar matter. We determine the location of the star surface, the event, and apparent horizons at a given proper time. It is seen that,  contrary to the Schwarzschild case,  it takes an infinite proper time for completing the collapse.\\ 
%Here, we have used three basic assumptions to simplify the calculations. First,
The outline of this paper is as follows: In section \ref{Gravitational collapse in a regular BH},  we consider the collapse of a non-radiating homogeneous star to a general regular BH and develop the standard OS formulation of gravitational collapse to it. Application of this description to the Hayward and Bardeen BHs is discussed in sections \ref{Gravitational collapse in Hayward BH} and \ref{Gravitational collapse in Bardeen BH} respectively. The evolution of the event and apparent horizons are studied and the validity of energy conditions for stellar matter is discussed. For both cases,  we find that the star is described by the polytropic equation of state albeit with a different polytropic index.\\
Instead of assuming that the metric outside the star is given by a regular BH, one can assume that the collapsing star consists of some polytropic matter with an arbitrary index  and find the exterior metric. In this way, analytically, we have shown that the final product of the gravitational collapse is a general static regular BH specified by two parameters of polytropic star. This is what we have discussed in section \ref{Gravitational collapse of a polytropic star to a general regular BH}. In the last section, we will provide a summary and present further discussions. Throughout this paper, the signature of the metric tensor is assumed to be (-, +, +, +). We use geometrized units, i.e., $G = c = 1$. A dimensionless variable is denoted with a tilde and obtained by normalizing it with respect to the Schwarzschild radius and a dot (a prime) over a  variable denotes its derivative with respect to  the proper time (radial coordinate) of a freely falling particle on the star surface. Here, the normalized initial radius of the star is set to $\tilde{R}_0=R_0/2m=2$.
\section{Gravitational collapsing into  regular BH}
\label{Gravitational collapse in a regular BH}
Here, we consider a particular ansatz for the form of a static, spherically symmetric and asymptotically flat regular BH metric as following 
\begin{align}
\label{0}
ds^2 = - \left(1 - \frac{2m}{r\alpha(r)}\right) dt^2 +  \left(1 - \frac{2m}{r\alpha(r)}\right)^{-1} dr^2 + r^2 d \Omega^2
\end{align}
where  $m$ is the geometrical mass of the star with the length dimension in the geometrized units. $\alpha(r)$ is an arbitrary dimensionless positive function that regularizes the singularity of Schwarzschild metric at r = 0 s. 
%\textbf{$\alpha \neq Constant$ represents matter distribtion outside of the star. In other words regular black holes have non-zero energy momentum tensor %everywhere except in infinity where the spacetime becomes flat, so one can deduce Birkhoff theorem prevails in infinity.}
Inserting (\ref{0}) into the Einstein equations, one obtains the density and anisotropic pressures of matter content 
\[
\rho^{(reg)}=-p_r^{(reg)}=-\frac{1}{4\pi}\frac{m\alpha'}{r^2\alpha^2}
\]
\begin{equation}
\label{pre}
p_\theta^{(reg)}=p_\varphi^{(reg)}=\frac{1}{8\pi}\frac{m(\alpha\alpha''-2\alpha'^2)}{r\alpha^3}
\end{equation}
filling all space and superscript (reg) denotes regular BH metric. For $\alpha (r)= \textit{cons.}$, equation (\ref{pre}) shows that the density and anisotropic pressure become zero, thus the Schwarzschild exterior solution is recovered. For  $\alpha (r) \neq \textit{cons.}$, metric (\ref{0}) is not the standard Schwarzschild one, however it does not contradict the Birkhoff's theorem since the region outside the compact object is not vacuum.\\
Considering regular BH spacetime (\ref{0}), here we will attempt to describe the gravitational collapse of a star according to the OS model \cite{6Mal}. In the standard OS scenario, the stellar matter is described by a homogeneous and isotropic  perfect fluid. Depending on the initial condition of star, the interior geometry is governed by a spatially closed or flat Friedmann equation and the exterior geometry is Schwarzschild BH. Moreover, the core geometry is joined smoothly to the BH spacetime and there is no need to assume a thin shell distribution of matter.

Here, we consider a compact star embedded in the regular spacetime (\ref{0}). The star does not reside in vacuum, therefore it presents some region  that is much denser than its surroundings, and so collapses. The stress--energy tensor of stellar matter occupying this region is unknown. However, it is straightforward to determine it by demanding a smooth transition of the metric across the boundary of star. 
Here, we consider the spatially flat FRW metric for the interior geometry of star and choose the time coordinate of the exterior (interior) metric to be the proper time of freely falling (comoving) particles. \\
It is convenient to write the regular metric (\ref{0}) in terms of  Painlev\'e-Gullstrand coordinates \cite{PG1,PG2,Mart} which are adopted to the freely falling radial observer starting from rest at infinity. This observer has the four-velocity $u^\alpha\partial_{\alpha}=(1 - 2m/{r\alpha(r)})^{-1}\partial_{t}-\sqrt{2m/{r\alpha(r)}}\partial_{r}$. By calculating $u_\alpha$, one can deduce the proper time of this observer by integrating  $d\tau =dt+(1 - 2m/{r\alpha(r)})^{-1}\sqrt{2m/{r\alpha(r)}}dr$. Inserting this expression for $dt$ into  (\ref{0}), the Painlev\'e-Gullstrand form of metric (\ref{0}) is obtained
\begin{align}
\label{6}
ds^2 = - d\tau^2 + (dr+ \sqrt{\frac{2m}{r\alpha(r)}} d \tau)^2 + r^2 d\Omega^2
\end{align}
%\begin{align}
%\label{6}
%ds^2 = - d\tau^2 + (dr+ \sqrt{\frac{2m}{r\alpha(r)}} d \tau)^2 + r^2 d\Omega^2
%\end{align}\\
%*****************
%Assuming that at the surface of the star, $r(\tau)=R(\tau)$, each freely falling particle starts from rest at infinity and  moves along a radial timelike %geodesic. The star surface is given by
%\begin{align}
%\label{6dot}
%\dot R(\tau) =-\sqrt{\frac{2m}{R\alpha(R)}}
%\end{align}
%*************
%Therefore, $H(\tau)$ can now be written as
%\begin{align}
%\label{2}
%H(\tau) = \frac{\dot R_0 (\tau)}{R_0 (\tau)} \Rightarrow \sqrt{R_0(\tau)^3 + 2m l^2} = \frac{\sqrt{2m}}{H(\tau)}
%\end{align}\\
In terms of Painlev\'e-Gullstrand cosmological coordinates, the spatially flat FRW interior geometry is expressed as 
\begin{align}
\label{9}
ds^2 = - d\tau^2 + (d r - r H(\tau) d \tau)^2 + r^2d\Omega^2
\end{align}
where $r(\tau)=a(\tau)r_c$, $H(\tau) = \dot{r}(\tau)/r(\tau)$, $a(\tau)$ is the scale factor and  $r_c$ is the comoving radial coordinate. 
%the spatially flat FRW interior geometry is expressed as 
%\begin{align}
%\label{9}
%ds^2 = - d\tau^2 + (d r - r H(\tau) d \tau)^2 + r^2d\Omega^2
%\end{align}
%where $r(\tau)=a(\tau)r_c$, $H(\tau) = \dot{r}(\tau)/r(\tau)$, $a(\tau)$ is the scale factor and  $r_c$ is the comoving radial coordinate
It is more convenient to write the interior metric (\ref{9}) as 
\begin{equation}
ds^2=-d\tau^2+a^2(\tau)(dr_c^2+r_c^2d\Omega^2)
\end{equation}
and use the exterior metric in the form of (\ref{0}). Following \cite{is}, a straightforward  calculation gives the components of the extrinsic curvature on the star surface, $r(\tau)=R(\tau)$ as
\[
{}^{(in)}K^{\tau}_{\tau}=0  \hspace{1cm} {}^{(in)}K^{\theta}_{\theta}={}^{(in)}K^{\varphi}_{\varphi}=1/R
\]
\begin{equation}
{}^{(out)}K^{\tau}_{\tau}=\dot{\gamma}/R \hspace{1cm} {}^{(out)}K^{\theta}_{\theta}={}^{(out)}K^{\varphi}_{\varphi}=\gamma/R
\end{equation}
where $\gamma=\sqrt{\dot{R}^2+1-2m/(R\alpha)}$. To get a smooth joining of two geometries, one must have $\gamma=1$ which yields to the following relation for the star surface
\begin{align}
\label{6dot}
\dot R(\tau) =-\sqrt{\frac{2m}{R\alpha(R)}}
\end{align}
This shows that at the surface of the star, each freely falling particle starts from rest at infinity and  moves along a radial timelike geodesic.
Therefore according to the Israel junction conditions, no energy layer appears on the star surface and there is a smooth transition across it. \\
Substituting $R(\tau)=a(\tau) R_c$ into the Freidmann equation 
\begin{equation} 
\label{f}
\dot{a}^2=\frac{8\pi}{3}\rho a^2
\end{equation}
%Where $\rho$ represents the density of matter content given by EMT on the surface of the star.} 
and comparing with (\ref{6dot}), one finds that the metric is the same  on both sides of the star if 
\begin{equation}
\label{rp}
\rho(\tau)=\frac{3m}{4\pi R^3(\tau)\alpha(R(\tau))}
\end{equation}
Taking the time derivative of (\ref{rp}) and substituting it into the continuity equation of the total perfect fluid filling the interior of the star,  $\dot \rho+3\dot a/a(\rho+p)=0$, gives the stellar surface pressure as
\begin{equation}
\label{p}
p(\tau)=\frac{m \alpha'}{4\pi R^2(\tau) \alpha^2(R(\tau))}
\end{equation}
The density (\ref{rp}) and the isotropic pressure (\ref{p}) represent the total energy density and pressure on the surface of star which are consisting of the background part  given by (\ref{pre}) and an unknown stellar part.\\
As mentioned before, we have assumed that any test particle on the star surface is a freely falling particle and hence its radial trajectory is obeying (\ref{6dot}). To justify this, using (\ref{pre}) and (\ref{p}), one can  easily calculate the pure radial pressure on the star surface as   $p-\left. p^{(reg)}_r\right|_R=0$. 
As expected, we see that the boundary surface of star can be specified by null radial pressure and so each particle follows the radial geodesic which is in confirmation with (\ref{6dot}).\\
Knowing $\alpha(R)$, one can find the stellar equation of state making use of (\ref{rp}) and (\ref{p}).
Moreover, from equation (\ref{rp}) it follows that
\begin{align}
\label{12.4}
m = \frac{4\pi}{3} \rho(\tau) R(\tau)^3 \alpha(\tau)=m_s(\tau)\alpha(\tau)
\end{align}
where $m_s$ gives the total mass of star.  As expected, in the Schwarzschild case, $\alpha=1$,  the total mass of star and the mass parameter of metric are the same.\\
%By a simple calculation, one can show that (\ref{12.4}) provides that the extrinsic curvature is the same on both sides of the star. 
As two illustrative examples, we consider the generalized OS gravitational collapse of a star to the two well-known regular BHs with deSitter core, Hayward and Bardeen, in the following sections.\footnote{A first attempt for fixing $\alpha$-function might obtain by setting $p_\theta^{(reg)}=p_\varphi^{(reg)}=0$ in (\ref{pre}) which leads to $\alpha (r)=(a+br)^{-1}$ with $a$ and $b$ being constants. Substituting it in the line element (\ref{0}), one can find that it is a singular BH. Performing an analysis similar to what is presented in the following sections, we obtain the standard OS gravitational collapse.}
\section{Gravitational collapsing into Hayward BH}
\label{Gravitational collapse in Hayward BH}
\begin{figure}[h]
\centering
\includegraphics[width=3in]{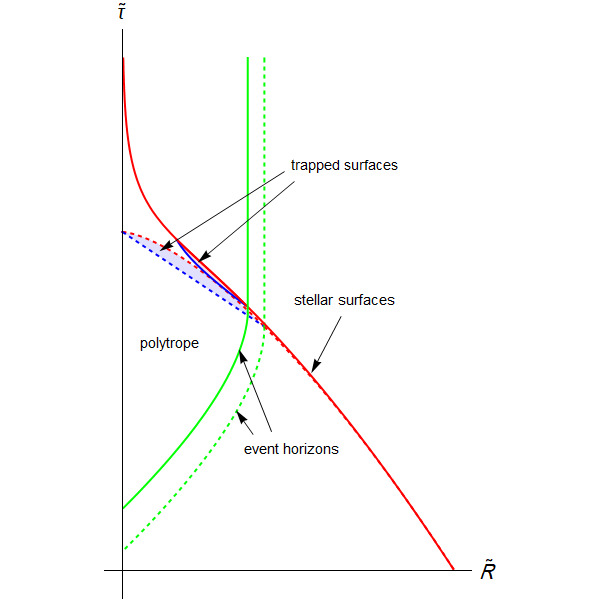}
\caption{The evolution of the star surface and horizons in the collapse of polytropic sphere to Hayward BH, setting $\tilde{l}=0.3$. Dashed and solid lines represent the OS collapse and its generalization for the Hayward BH respectively.}
\label{figB1}
\end{figure}

One of the simplest regular model of a BH proposed by Hayward in \cite{54Fro16} for which 
\begin{equation}
\label{aph}
\alpha(r) = 1 + 2 m l^2/r^3
\end{equation}
where $l$ is a length scale parameter. It follows from (\ref{aph}) that the metric (\ref{0}) reduces to the Schwarzschild and deSitter one in the limit $r \to \infty$ and $r \to 0$ respectively. Moreover, this metric is a solution of Einstein equations with the following anisotropic fluid as source
\[
\rho^{(H)}=-p_r^{(H)}=\frac{1}{8\pi}\frac{12l^2m^2}{(r^3+2ml^2)^2}
\]
\begin{equation}
\label{preh}
p_\theta^{(H)}=p_\varphi^{(H)}=\frac{1}{8\pi}\frac{24l^2m^2(r^3-ml^2)}{(r^3+2ml^2)^3}
\end{equation}
Substituting (\ref{aph}) in (\ref{6dot}), one finds that 
\begin{align}
\label{6dot2}
\dot R(\tau) =-\frac{ \sqrt{2m}R}{\sqrt{R^3 + 2ml^2}}
\end{align}
which can be expressed in terms of dimensionless variables as 
\begin{equation}\label{dim}
\dot{\tilde{R}}(\tilde{\tau})=-\frac{\tilde{R}}{\sqrt{\tilde{R}^3+\tilde{l}^2}}
\end{equation}
 and gives
\[
\frac{3}{2}\tilde\tau=\left(\sqrt{\tilde{R}_0 ^3 +\tilde{l}^2}-\sqrt{\tilde{R}^3 +\tilde{l}^2}\right)+
\]
\begin{align}
\label{7}
\tilde{l}\left(\tanh^{-1}\sqrt{1+\frac{\tilde{R}^3}{\tilde{l}^2}}-\tanh^{-1}\sqrt{1+\frac{\tilde{R}_0^3}{\tilde{l}^2}}\right)
\end{align}
where the integration constant $\tilde{R}_0$ is chosen such that $\tilde{\tau}=0$ at $\tilde{R}=\tilde{R_0}$. Clearly in the case of $l = 0$,  equation (\ref{6dot2}) reduces to the corresponding one for the OS collapse i.e. $\tilde R^{1/2}  \dot{\tilde R} = - 1$ and after an integration, gives  $3\tilde{\tau}/2=\tilde{R_0}^{3/2}-\tilde{R}^{3/2}$ (See equation (29.56) of \cite{Blau}). By expanding (\ref{7}), one finds that at the late stages of collapse 
$\tilde{R}\sim e^{-\tilde{\tau}/{\tilde{l}}}$, therefore the rate of collapse decreases with increasing $\tilde{l}$.  Setting $\tilde{l}=0.3$, the result (\ref{7})
is plotted in figure \ref{figB1}. It is clear that it takes an infinite proper time for the star to contract completely. Also, for small values of $\tilde{R}$, the star radius decreases more slower than that of the standard OS \cite{rez}. Other details of this figure are discussed in the next subsection.
% For the dynamics of different surfaces in the OS collapse see }
% and also one can find that it decreases with increasing $\tilde{l}$.

One can use (\ref{aph}) to express the density and pressure  (\ref{rp}) and (\ref{p}) as a function of $R$ which itself is a function of the proper time according to (\ref{7}). 
Eliminating  $R$ in the resulted expressions leads to the equation of state of star.  The result is as follows
\begin{equation}
\label{plat}
p(\tau)=-\frac{8\pi}{3}l^2\rho^2(\tau) \hspace{1cm} \rho(\tau)=\frac{3m}{4\pi(R(\tau)^3+2ml^2)}
\end{equation}
Therefore, in the case of Hayward BH, the metric is continuous across the star surface if the stellar matter described by a polytropic 
fluid, $p \propto \rho^{1+1/n}$ in which the polytropic index, $n$, equals to unity. It is worth mentioning that many gaseous planets and
stars \cite{Kong1} and also, the dynamical evolution of the rotating Bose–Einstein condensate dark matter halos \cite{Harko} can be approximately described by a polytropic equation of state with index unity. \\
The negative pressure in (\ref{plat})  acts as an anti-gravitational effect which resists against the gravitational collapse \cite{med}. It is known that avoiding the singularity at the end times of collapsing process needs exotic matter with negative pressure. This kind of matter appears in Gravastar model \cite{25Mal}, Schwarzschild-Ads BH with a minimal length \cite{Yan} and other non-singular BHs resulted from gravitational collapse \cite{ddMb}.
The negative pressure indicates that demanding OS collapse in Hayward spacetime may lead to the violation of some energy conditions.
%\textbf{The negative pressure in  (\ref{plat})  is limited inside of the star which results in avoiding the singularity. One may ponder about the physical %description of the model, the negative pressure do not bring the model into question. The regular black hole models like Hayward reduce to de Sitter spacetime, %which is one of the main models in cosmology where dark energy with negative pressure is dominant. Also, the negative pressure indicates that demanding OS %collapse in Hayward spacetime may lead to the violation of some energy conditions. It is straightforward to apply these conditions to the stellar matter %described by (\ref{plat}).}
Assuming a positive energy density, the dominant energy condition (DEC)($\rho\geq|p|$) leads to $\rho\leq 3/8\pi l^2$ which holds trivially by substituting (\ref{plat}). Therefore the weak ($\rho\geq0$, $\rho+p>0$) and null energy ($\rho+p\geq0$) conditions are also satisfied inside the star. However, in terms of the dimensionless variables, the SEC ($\rho+3p\geq0$, $\rho+p\geq0$) is violated for
\begin{equation}\label{SEC}
\tilde{R}(\tau)\leq(2\tilde{l}^2)^{1/3}
\end{equation}
Therefore,  as expected, when the normalized radius of  star becomes smaller than $(2\tilde{l}^2)^{1/3}$, one can say that the stellar distribution involves some ab-normal matter \cite{Viss}. This is the key feature of a regular BH for which the SEC is violated at a small radius where the metric becomes de Sitter like \cite{40FroPrd16,silvajadid}. 
\\
Let us now attempt to investigate the gravitational mass of the collapsing star. This can be immediately obtained from the relations (\ref{12.4}) and (\ref{aph}) as
\begin{align}
\label{12dot4}
m = \frac{4\pi}{3} \rho R^3 + \frac{8\pi}{3} \rho m l^2
\end{align}
The first term on the right-hand side gives the total mass of star, $m_s$. We may therefore write (\ref{12dot4}) in terms of  dimensionless parameters as 
\begin{equation}\label{mm}
\frac{m}{m_s(\tau)}=1+\frac{\tilde{l}^2}{\tilde{R}^3(\tau)}
\end{equation}
We see that the length scale parameter of Hayward BH reduces the total mass of star.\\
%A straightforward calculation, using (\ref{preh}) and (\ref{plat} shows that \textbf{the pure radial pressure on the star surface  $p_r-p^{(reg)}_r=0$ as %expected.}
%\[
%\rho^{(S)}= \frac{3mR(\tau)
%^3}{4\pi(R(\tau)^3+2ml^2)^2}
%\] 
%\begin{align}
%\label{pointone}
%p^{(S)}_r=0  \ \ \ \ \ p^{(S)}_{\theta}=p^{(S)}_{\varphi}= -\frac{9l^2m^2R(\tau)^3}{2\pi(R(\tau)^3+2ml^2)^3}
%\end{align}
By obtaining $\tilde{H}(\tilde\tau)=\dot{\tilde{R}}(\tilde{\tau})/\tilde{R}(\tilde{\tau})$ from (\ref{dim}) and then substituting it   into (\ref{9}), one can easily see that the form of interior metric is the same as (\ref{0}) at the surface of star.
Moreover, the evolution of $\tilde{H}(\tilde{\tau})$ can be obtained from (\ref{dim}) and (\ref{7}) as follows
\begin{align}
\label{10}
\frac{3}{2}\tilde{\tau}=\left(\frac{1}{\tilde{H}}-\frac{1}{\tilde{H}_0}\right)+\tilde{l}\left(\tanh^{-1}\frac{1}{\tilde{H}_0 \tilde{l}}-\tanh^{-1}\frac{1}{\tilde{H}\tilde{l}}\right)
\end{align}
where $\tilde{H}_0=-1/\sqrt{\tilde{R}_0 ^3 +\tilde{l}^2}$ according to (\ref{6dot2}). One may compare (\ref{10}) with $3\tilde{\tau}/2=1/\tilde{H}-1/\tilde{H}_0$ resulted directly from (\ref{6dot2}) for  a Schwarzschild BH \cite{Blau}. 

\subsection{Horizons}
\label{Horizons}

We now consider the evolution of the  event and apparent horizons. As mentioned before, the exterior and interior geometry is given by equations (\ref{6}), (\ref{9}) and (\ref{aph})  where the Hubble parameter is negative. 

\begin{figure}[h]
\centering
\includegraphics[width=3in]{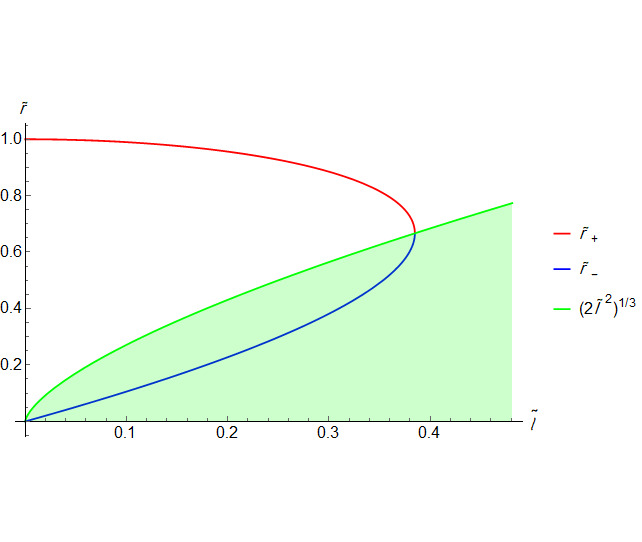}
\caption{$\tilde{r}_+$ and $\tilde{r}_-$ with respect to $\tilde{l}$. In the shaded region, the SEC is violated.}
\label{fig1-1}
\end{figure}

In the exterior region of collapsing star, the event and apparent horizons are the same and can be found by solving the following third order equation
\begin{equation}\label{hor}
r^3-2m r^2+2m l^2=0
\end{equation}
It has three real roots  if $27l^2<16 m^2$ or equivalently $\tilde{l}^2<4/27$. In this case, two roots are positive \cite{5} 
\begin{align}
\label{18}
\tilde{r}_{\pm}=\frac{1}{3} +\frac{2}{3} \cos\left(\frac{\pi}{3} \mp \frac{1}{3}\cos^{-1} (\frac{27\tilde{l}^2}{2}-1)\right)
\end{align}
in which $\tilde r_{+}$ and $\tilde r_{-}$ are outer and inner horizons. In figure \ref{fig1-1} we have plotted the locations of horizons with respect to $\tilde{l}$ and the region specified by (\ref{SEC}) \cite{5}. The upper and lower curves show $\tilde r_{+}$ and $\tilde r_{-}$ respectively.  In this figure it can also be seen nicely which values of $\tilde{l}$ correspond to a BH with two horizons, an extremal BH and a horizon-less compact object. According to (\ref{SEC}),  the SEC is violated in the shaded region which is located inside the outer horizon.

Assuming that the star crosses $\tilde{r}_{+}$ at $\tilde{\tau}_f$, i.e. $\tilde{R}(\tilde{\tau}_f)=\tilde{r}_{+}$, therefore from (\ref{7}), we find that
\begin{align}
\label{77}
\frac{3}{2}\tilde{\tau_f}&=\left(\sqrt{\tilde{R}_0^3 + \tilde{l}^2}-\tilde{r}_{+}\right)+ \nonumber\\
&+\tilde{l}\left(\tanh^{-1}\frac{\tilde{r}_{+}}{\tilde{l}}-\tanh^{-1}\sqrt{1+\frac{\tilde{R}_0^3}{\tilde{l}^2}}\right) 
\end{align}
where the relation $\tilde{r}_{+}^3+\tilde{l}^2=\tilde{r}_{+}^2$  is used according to (\ref{hor}).
In the interior of star, noting (\ref{9}), the radial null geodesics are given by $dr/d\tau = - 1 + r H$ and $dr/d\tau = (1 + r H )$. Since $H<0$, the first one corresponds to the ingoing null geodesics while the second represents the ingoing and outgoing null geodesics for $r>-1/H$ and $r<-1/H$ respectively.
This means that there are some trapped surfaces whose radii are smaller than $-1/H$ and thus the dimensionless interior apparent horizon is given by  $\tilde{r}_{ah}=-1/\tilde{H}$. Developing an apparent horizon inside the star means that a BH is formed. Substituting $\tilde{r}_{ah}=-1/\tilde{H}$  into equation (\ref{10}), we obtain 
\begin{align}
\label{ah}
\frac{3}{2}\tilde{\tau}={-(\tilde{r}_{ah}}+\frac{1}{\tilde{H}_0})+\tilde{l}(\tanh^{-1}{\frac{\tilde{r}_{ah}}{\tilde{l}}}+\tanh^{-1}\frac{1}{\tilde{l}\tilde{H}_0})
\end{align}
in which $-1/\tilde{H}_0$ is indeed $\tilde{r}_{ah}$ at $\tilde{\tau}=0$. A plot of $\tilde r_{ah}$ as a function of $\tilde{\tau}$ is shown in figure \ref{figB1}. We see that the trapped surfaces form when the star surface crosses $r_+$ and unlike the OS collapse, the apparent horizon doesn't go to zero finally.
%approaches $\tilde{l}$ at the end stage of collapse. 
%To find the value of this constant, we can expand (\ref{ah}) as $\tilde r_{ah}\sim\tilde{l}(1+ c\ e^{- \tilde{\tau}/\tilde{l}})$
% where $c$ is a constant depending on $ \tilde{R}_0$ and $\tilde{l}$. Therefore, the normalized apparent horizon tends to $\tilde{l}$  finally.
Comparing (\ref{77}) with (\ref{ah}), a simple calculation shows that $\tilde{r}_{ah}(\tilde{\tau}_f)=\tilde{r}_+$, as expected. 
It is also worth remembering that for the standard OS collapse where $\tilde{r}_+=1$ and $\tilde{H}(\tau_f)=-1$, equations (\ref{77}) and (\ref{ah}) reduce to $3\tilde{\tau}_f /2 = \tilde{R}_0 ^{3/2}-1$ and  $3\tilde{\tau} /2 =1/\tilde{H}_0 -\tilde{r}_{ah}$ \cite{Blau}.\\
It may be useful here to mention that the normal vector of the apparent horizon (\ref{ah}) satisfies
\begin{align}
\label{16}
n_\alpha n^\alpha \varpropto \frac{\tilde r_{ah}^2+3\tilde{l}^2}{\tilde r_{ah}^2-\tilde{l}^2}
\end{align}
for which we have used the inverse of metric (\ref{9}) evaluated on the apparent horizon. 
According to (\ref{6dot2}), $\tilde{r}_{ah}=\sqrt{\tilde{R}^3+\tilde{l}^2}$, so that $\tilde{r}_{ah} > \tilde{l}$ and therefore, during the gravitational collapse, the  apparent horizon is a timelike surface.

To determine the interior event horizon, $r_{eh}$, we consider the outgoing null geodesic arriving the surface of star at $\tau_f$ , i.e.,  $R(\tau_f)=r_+$. As mentioned before, the outgoing null geodesics satisfy
\begin{equation}\label{r}
\dot {\tilde r}_{eh} = 1 + \tilde{r}_{eh}\tilde{H}
\end{equation}
By virtue of (\ref{6dot2}) and writing $\dot {\tilde r}_{eh}= (d {\tilde r}_{eh}/d\tilde{R})\dot{\tilde{R}}$, equation (\ref{r}) reads
\begin{equation}
\frac{d\tilde{r}_{eh}}{d\tilde{R}}-\frac{\tilde{r}_{eh}}{\tilde{R}}=-\frac{\sqrt{\tilde{R} ^3 +\tilde{l}^2}}{\tilde{R}^2}
\end{equation}
Remembering that $\tilde{{r}}_{eh} (\tilde{r}_+) = \tilde{r}_+$, the above equation can be solved, and gives 
\[
\tilde{r}_{eh}=\tilde{R} +
\]
\begin{align}
\label{17}
\tilde{l} \left({}_2 F_1 [\frac{-1}{2}, \frac{-1}{3}, \frac{2}{3},-\frac{\tilde{R}^3}{\tilde{l}^2}]
-\frac{\tilde{R}}{\tilde{r}_+} {}_2 F_1 [\frac{-1}{2}, \frac{-1}{3}, \frac{2}{3},-\frac{\tilde{r}_+ ^3}{\tilde{l}^2}]\right)
\end{align}
where ${}_2 F_1$ is the hypergeometric function. Equations
 (\ref{10}) and (\ref{17}) give the evolution of the event horizon as a function of the proper time. The result is shown in figure \ref{figB1}.  Such as the OS collapse, the event horizon starts from zero radius and then reaches $r_+$ when the star surface is $r_+$. Following similar calculation as above for the standard OS collapse leads to $\tilde{r}_{eh}=-2\tilde{R}^{3/2}+3\tilde{R}$ \cite{Blau} where as mentioned before, $\tilde{R}=(\tilde{R_0}^{3/2}-3\tilde{\tau}/2)^{2/3}$. 
%From figure \ref{figB1}, we see that the event horizon has a local maximum and then goes to $\tilde{l}$ at sufficient large values of the proper time.
%\textbf{ which is the end of the collapse and never reaches zero.}

\section{Gravitational collapsing into Bardeen BH}
\label{Gravitational collapse in Bardeen BH}
\begin{figure}[h]
\centering
\includegraphics[width=3in]{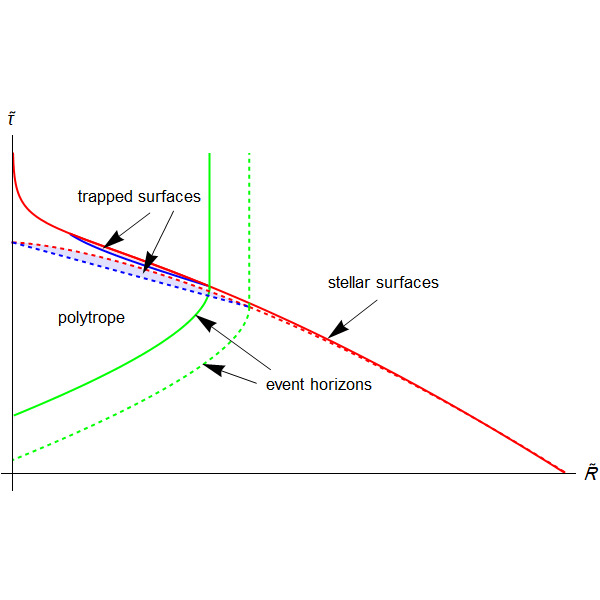}
\caption{The evolution of the star surface and horizons in the collapse of polytropic sphere to Bardeen BH, setting $\tilde{g}=0.3$. Dashed and solid lines represent the OS collapse and its generalization for the Bardeen BH respectively.}
\label{figBRR}
\end{figure}

Similar calculations can be done for Bardeen BH \cite{3RodJcap-9-21Kho} which is the first singularity-free BH resulted from coupling of Einsteinian gravity to a non-linear electrodynamics field. The Bardeen metric is given by (\ref{0}), where
\begin{equation}
\label{alpa3}
\alpha = (1+g^2/r^2)^{3/2}
\end{equation}
and $g$ is the magnetic charge with length dimension in the geometrized units.
The matter density and anisotropic pressures associated with this metric are \cite{silva}
\begin{align}
\label{silva}
\rho^{(B)} &= -p_r^{(B)} = \frac{1}{8\pi}\frac{6 m g^2}{( r^2 + g^2)^{5/2}}\nonumber \\
p_\theta^{(B)} &= p_\phi^{(B)} = \frac{1}{8\pi}\frac{3 g^2 m (3 r^2 - 2 g^2)}{(g^2+r^2)^{7/2}}
\end{align}
Substituting (\ref{alpa3}) in (\ref{6dot}), one finds that 
\begin{align}
\label{0-8}
\dot R(\tau) = - \frac{\sqrt{2m} R}{(R^2 + g^2)^{3/4}}
\end{align}
Introducing the dimensionless variable $\tilde{g}$, (\ref{0-8}) reduces to 
\begin{align}
\label{00-88}
\dot{\tilde{R}}(\tilde{\tau})=-\frac{\tilde{R}}{(\tilde{R}^3+\tilde{g}^2)^{3/4}}
\end{align}
which has the following solution
\[
\frac{3}{2}\tilde\tau = (\tilde g^2 + \tilde R_0^2)^{3/4} -(\tilde g^2 + \tilde R^2)^{3/4}+
\]
\[
\frac{3}{2}\tilde g^{3/2} \left(-\tan^{-1} \big( \frac{\tilde g^2+  \tilde R^2}{\tilde g^2}\big)^{1/4} +\tan^{-1} \big( \frac{\tilde g^2+  \tilde R_0^2}{\tilde g^2}\big)^{1/4}+ \right.
\]
\begin{align}
\label{0-9}
\left. \tanh^{-1} \big( \frac{\tilde g^2+ \tilde R^2}{\tilde g^2}\big)^{1/4}- \tanh^{-1} \big( \frac{\tilde g^2+ \tilde R_0^2}{\tilde g^2}\big)^{1/4}\right)
\end{align}
This solution exhibits behavior similar to that found in the Hayward case. For small radius of star
\begin{equation}
\tilde\tau =f(\tilde{R_0},\tilde{g})- \tilde{g}^{3/2}\ln{\tilde R}) - \frac{3\tilde{R}^2}{8\sqrt{\tilde{g}}} + O(\tilde{R}^3)
\end{equation}
Therefore at the final stage of collapse $\tilde R \sim e^{-\tilde{\tau}/\tilde{g}^{3/2}}$, so the parameter $\tilde{g}$ controls the rate of contraction at late times.  Moreover, the rate of decreasing of $\tilde R$ is more sensitive to the variation of $\tilde{g}$ than to $\tilde{l}$ in the Hayward case. In  figure \ref{figBRR} we have plotted the worldline of star surface by setting $\tilde g=0.3$. See the following subsection for other details of this figure.
%\textbf{It should be pointed that the radius never reaches zero, in other words, the observers needs infinite amount of time to reach the zero point, Meaning %the zero point of the black hole is unreachable.}

Substituting $R(\tau)=a(\tau)R_c$, into (\ref{f}) and comparing with (\ref{0-8}), one can find the equation of state of star 
\begin{align}
\label{0-92}
p(\tau) = -(\frac{4 \pi}{3 m})^{2/3} g^2 \rho(\tau)^{5/3} \ \ \ \ \ \rho(\tau) = \frac{3m}{4\pi(R(\tau)^2 + g^2)^{3/2}}
\end{align}
Therefore the collapsing star consists of a polytropic fluid with index $n=3/2$ which is usually used to model the convective star cores and also the low mass white dwarfs \cite{Sage}.
Assuming a positive energy density, a simple calculation shows that the DEC holds trivially and only the SEC is violated when the radius of star becomes 
\begin{equation}\label{SECB}
\tilde{R}<\sqrt{2}\tilde{g}
\end{equation}
%\textbf{Null and weak energy conditions are satisfied everywhere and dominant energy condition is satisfied where $R\geq\sqrt{1-g^2}$ is established.}
Moreover, as mentioned in the previous section, the extrinsic curvature is the same on both sides of the star surface. This is the straightforward result of (\ref{12.4}) and (\ref{alpa3}) which can be written as
\begin{align}
\label{m0-2}
\frac{m}{m_s(\tau)} = (1+ \frac{\tilde{g}^2}{\tilde{R}(\tau)^2})^{3/2}
\end{align}
Another quantity that we are interested in is 
%\[
%\rho^{(S)}= \frac{3mR(\tau)^2}{4\pi(R^2(\tau)+ g^2)^{5/2}}
%\]
%\begin{equation}
%p^{(S)}_r=0  \ \ \ \ \ p^{(S)}_{\theta}=p^{(S)}_{\varphi}= -\frac{15 R^2 g^2 m}{8\pi(R^2 + g^2)^{7/2}}
%\end{equation}
the dimensionless Hubble parameter which can be read from (\ref{0-8}) and (\ref{0-9}) as
\begin{align}
\label{0-11}
\frac{3}{2}\tilde{\tau} &=\big(\frac{1}{\tilde H} - \frac{1}{\tilde{H}_0}\big) +\frac{3}{2} \tilde{g}^{3/2} \big( \tan^{-1} \frac{1}{\tilde{H}^{1/3}\tilde{g}^{1/2}} \nonumber \\
&-\tan^{-1} \frac{1}{{\tilde{H}_0}^{1/3}\tilde{g}^{1/2}} -\tanh^{-1}\frac{1}{\tilde{H}^{1/3} \tilde{g}^{1/2}}\nonumber \\
&+\tanh^{-1}\frac{1}{{\tilde{H}_0}^{1/3} \tilde{g}^{1/2}}\big)
\end{align}
where the first two terms in the right hand side are the ones relevant to the standard OS collapse. 
\subsection{Horizons}
\begin{figure}[h]
\centering
\includegraphics[width=3in]{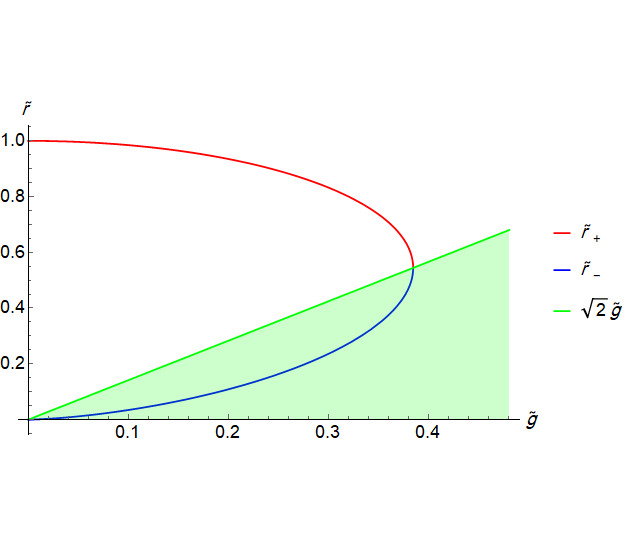}
\caption{${\tilde r}_+$ and ${\tilde r}_-$ as functions of $\tilde g$. In the shaded region the SEC is violated.}
\label{figBR}
\end{figure}

Outside the star, the event and apparent horizons coincide and can be found by solving 
\begin{align}
\label{BBH1}
\tilde{r}^2-\tilde{r}^{4/3}+\tilde{g}^2 =0
\end{align}
For $\tilde{g}^2\leq 4/27$, this equation has two positive roots
\begin{align}
\label{BBH2}
\tilde{r}_{\pm}=\left(\frac{1}{3} +\frac{2}{3} \cos (\frac{\pi}{3} \mp \frac{1}{3}\cos^{-1} (\frac{27\tilde{g}^2}{2}-1))\right)^{3/2}
\end{align}
%where $m_{-}=2$ and $m_{+}=0$. Also $a$, and $ {\delta}_{0 ,1}$ are defined as 
%\begin{align}
%\label{BBH3}
%a &= 1- 3 \tilde g^2\nonumber \\
%{\delta}_{0} &=\frac{2}{3} (1 - 6 \tilde g^2)^{1/2}\nonumber \\
%{\delta}_{1}&= \frac{1}{2} (1 - 6 \tilde g^2)^{-3/2} ( 2 - 18 \tilde g^2+ 27 \tilde g^4)
We have plotted ${\tilde{r}}_\pm$ in figure \ref{figBR}. The shaded region is where the SEC is violated. It can be seen that the overall behavior is similar to that found for Hayward BH.
% however, here the boundary of the region where strong energy condition is violated has slightly moved toward $ {\tilde{r}}_+$ in %the plane of ($\tilde{r}$, $\tilde{g}$).

The dimensionless apparent horizon, $\tilde{r}_{ah}$, can be read from equation (\ref{0-11}) as
\begin{align}
\label{0-12}
\frac{3}{2}\tilde{\tau} &=-\big(\tilde{r}_{ah} + \frac{1}{\tilde{H}_0}\big) +\frac{3}{2} \tilde{g}^{3/2} \big( - \tan^{-1} \frac{\tilde{r}_{ah}^{1/3}}{\tilde{g}^{1/2}} \nonumber \\
&-\tan^{-1} \frac{1}{{\tilde{H}_0}^{1/3}\tilde{g}^{1/2}} +\tanh^{-1}\frac{\tilde{r}_{ah}^{1/3}}{ \tilde{g}^{1/2}}\nonumber \\
&+\tanh^{-1}\frac{1}{{\tilde{H}_0}^{1/3} \tilde{g}^{1/2}}\big)
\end{align}
It is plotted in figure \ref{figBRR}. We see that in this case, the apparent horizon follows the same evolution as the Hayward one.
% a ( 1 + b e^{-12\tilde{\tau}/\tilde{g}^{3/2}})$ in which $b$ is a constant depends on $\tilde g$ and $\tilde{R}_0$.
The normal vector of the apparent horizon satisfies
\begin{align}
\label{0-13}
n_\alpha n^\alpha \varpropto \frac{\tilde{r}^{4/3}+3\tilde{g}^2}{\tilde{r}^{4/3}-\tilde{g}^2}
\end{align}
By noting that $\tilde{r}_{ah}=-1/\tilde{H}$, it is clear from (\ref{0-8}) that $\tilde r_{ah} > \tilde{g}^{3/2}$, therefore  the apparent horizon would be timelike during the collapse. \\
Now using the outgoing null geodesic equation and doing a calculation similar to the previous section, lead to the following expression for the event horizon
\[
\tilde{r}_{eh}=\tilde{R} +
\]
\begin{align}
\label{0-15}
\tilde{g}^{\frac{3}{2}} \left({}_2 F_1 [\frac{-3}{4}, \frac{-1}{2}, \frac{1}{2},-\frac{\tilde{R}^2}{\tilde{g}^2}]
-\frac{\tilde{R}}{\tilde{r}_+} {}_2 F_1 [\frac{-3}{4}, \frac{-1}{2}, \frac{1}{2},-\frac{\tilde{r}_+ ^2}{\tilde{g}^2}]\right)
\end{align}
Equations (\ref{0-9}) and (\ref{0-15}) determine the event horizon as a function of the proper time. It is plotted in figure \ref{figBRR}. We see that the evolution of event horizon is similar to that of Hayward BH discussed in the previous section.

%a ( 1 + b^\prime e^{-12\tilde{\tau}/\tilde{g}^{3/2}})$ where $b^\prime$ is a constant. By increasing $\tilde \tau$,  the event horizon finally %approaches to $\tilde{g}^{3/2}$.

\section{Gravitational collapsing of a polytropic star into general regular BH}
\label{Gravitational collapse of a polytropic star to a general regular BH}
\label{ap}
\begin{figure*}[h]
\centering
\includegraphics[width=7in]{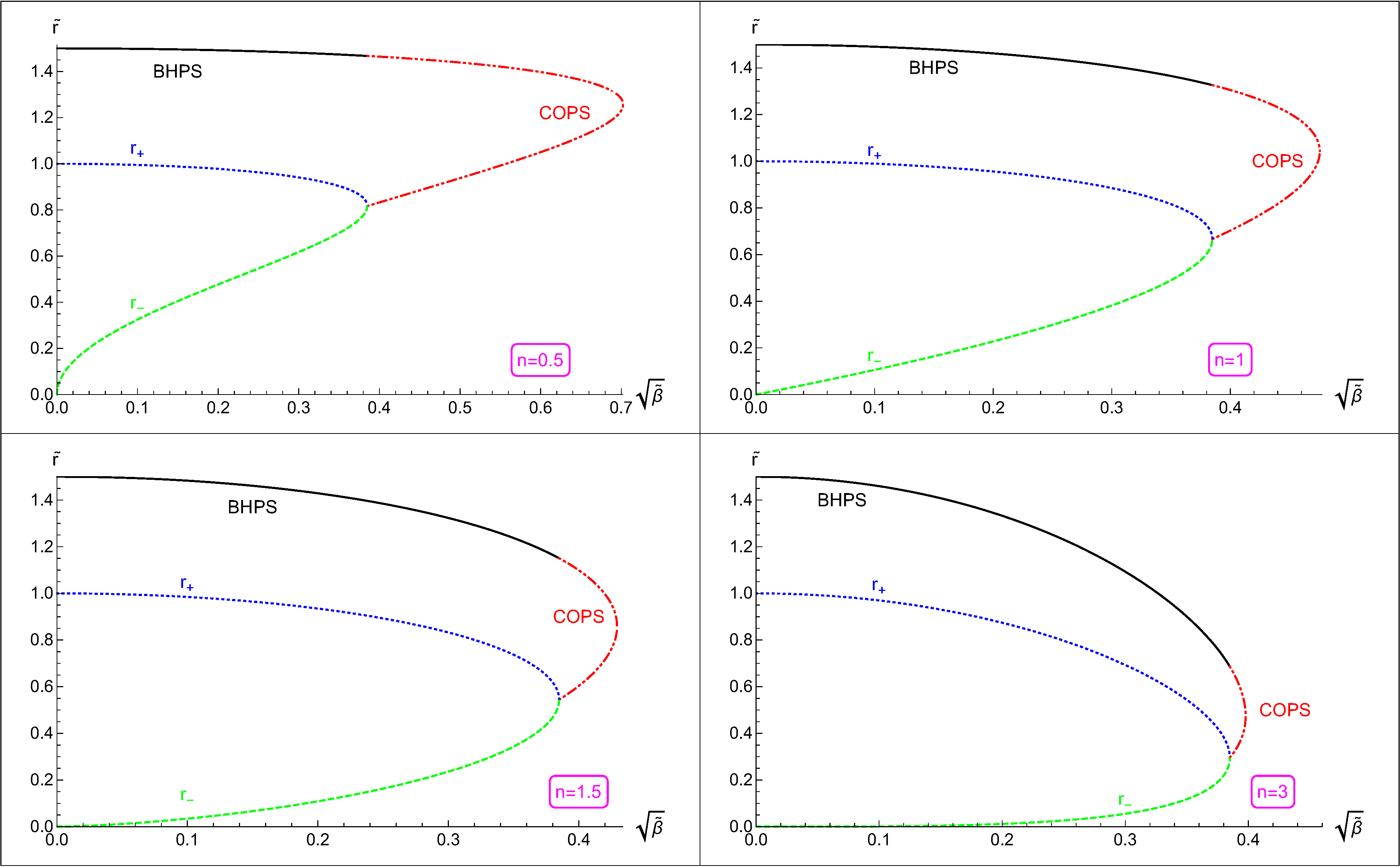}
\caption{The photon sphere and the inner and outer horizons as a function of $\sqrt{\tilde \beta}$. The black line is the BH photon sphere, BHPS. The dahsed-dotted red line is the photon spheres of horizonless compact object, COPS. The dotted blue and dashed green lines are $\tilde{r}_+$ and $\tilde{r}_-$ respectively.}
\label{photon}
\end{figure*}

In this section, we will do a kind of reverse procedure of the previous two sections. We seek a new family of spherically symmetric and static regular BHs in the form of (\ref{0}) in which there is a collapsing star governed by the polytropic equation of state and moreover, the star surface is not an energy layer, i.e. the interior metric of star jointed smoothly to the exterior regular BH solution.  We assume that the metric inside the star is described by a spatially flat FRW metric. 
%\textbf{The physical requirements for a general $\alpha$ is first, the chosen $\alpha$ must reduce to Hayward and Bardeen is a specified limit. Second, we %require the model to smoothly tend to zero radius and not reach zero and third, the model should follow the pressure and density expressions explained in %section \ref{Gravitational collapse in a regular BH}. Recalling (\ref{pre}), the case which  $p_\theta^{(reg)}=p_\varphi^{(reg)}= 0$ leads us 
%to $ \alpha = 1/(a - b r)$ causing the radius not to tend zero smoothly which violates the the second rule, which means the expression (\ref{pre}) must be a %non-zero quantity. Considering the mentioned requirements and 
Combining the general forms of the density and pressure, (\ref{rp}) and (\ref{p}), with the polytropic equation of state, $\tilde p \varpropto \tilde\rho^{1+1/n}$, one obtains
\begin{equation}
\label{alph}
\alpha= (1+\tilde\beta \tilde r^{-3/n})^n
\end{equation}
where $\tilde{\beta}$ is an arbitrary dimensionless positive parameter and we assume that the polytropic index $n$ is positive.
% \textbf{and not limited to a specific dimension as we have seen in Hayward and Bardeen, it can be scale parameter or magnetic charge. The EMT specifies the %nature of the $\beta$ and its dimension.}
The above relation also gives the total mass of star according to (\ref{12.4}). Putting (\ref{alph}) into (\ref{0}), we find a new family of regular BH metrics specified by two parameters, $\tilde\beta$ and $n$
\begin{align}
\label{end0}
\tilde {ds}^2 &= - (1 - \frac{1}{\tilde r (1+ \tilde\beta \tilde r^{-3/n})^n}) \tilde{dt}^2 +\nonumber \\
&(1 - \frac{1}{\tilde r (1+ \tilde\beta \tilde r^{-3/n})^n})^{-1} d\tilde r^2 + \tilde r^2 d \Omega^2
\end{align}
for the exterior region of the collapsing star. For sufficiently small (large) radius, (\ref{end0}) reduces to the deSitter (Schwarzschild) metric where $1/{\tilde \beta^n}$  plays the role of cosmological constant. Metric (\ref{end0}) satisfies Einstein equations provided that 
\begin{align}
 \rho^{(G)}&=-  p_r^{(G)}= \frac{1}{8\pi}\frac{3\tilde{\beta}}{(\tilde{\beta}+\tilde{r}^{3/n})^{n+1}}\nonumber \\
 p_\theta^{(G)}&= p_\varphi^{(G)}= \frac{1}{16\pi} \frac{ 3\tilde\beta \left((n+3){\tilde r}^{3/n}-2n\tilde\beta\right)}{(\tilde{\beta}+\tilde{r}^{3/n})^{n+2}}
\end{align}
Before discussing the gravitational collapse of a star into the BH space-time (\ref{end0}), it is instructive to consider some physically observable quantities. Herein, we focus on photon spheres and discuss their modifications by the line element (\ref{end0}). The effective potential for null trajectories is: 
\begin{equation}\label{VV}
\tilde{V}=\frac{1}{\tilde{r}^2}\left(1 - \frac{1}{\tilde r\alpha(\tilde r)}\right)
\end{equation}
where it is scaled with the angular momentum of photon and $\alpha$ is given by (\ref{alph}). The potential (\ref{VV}) tends to $+
\infty$ as $\tilde r\rightarrow 0$ and approaches zero at $\tilde r\rightarrow +\infty$. Thus in general, it has two extremal points. The smaller one presents a stable photon sphere while the other is unstable. The location of these orbits together with the outer and inner horizons, ${\tilde r}_+$ and ${\tilde r}_ -$  is plotted in figure \ref{photon} as a function of $\sqrt{\tilde \beta}$ for $n=0.5, n=1$ (Hayward BH), $n=1.5$ (Bardeen BH) and $n=3$. In each plot, the unstable BH photon sphere, BHPS, is presented. One can see that depending on $n$, at a specific value of $\tilde \beta$, the two horizons come together. For larger values of it, one is dealing with a horizonless compact object with two photon spheres, COPS. It should be noted that the orbit of circular photon is terminated at the horizon because the curve is not null after this point.\footnote{For a detailed analysis of photon spheres and the orbits of massive particles for a specific model of regular BH see \cite{univ}.}

\begin{figure}
\centering
\includegraphics[width=3in]{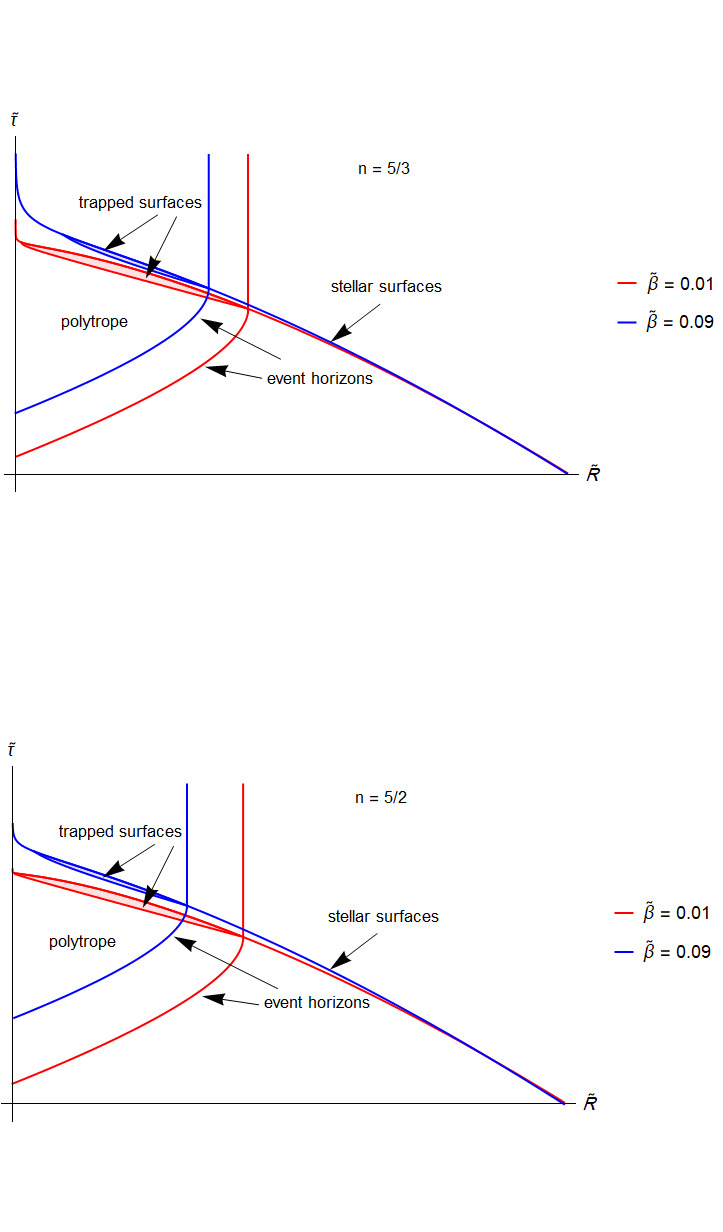}
\caption{The evolution of the star surface and horizons in the collapse of polytropic sphere to a general regular BH, setting $n=5/3$ and $\tilde\beta=0.01,0.09$ (up panel) and $n=5/2$ and $\tilde\beta=0.01,0.09$ (down panel). Dashed and solid lines represent the OS collapse and its generalization for the regular BH (\ref{end0}) respectively.}
\label{nplot1}
\end{figure}

Moreover, relations (\ref{rp}) and (\ref{p}) imply that 
\begin{align}
\label{genp}
\tilde\rho(\tilde\tau)=\frac{3}{8\pi}\frac{1}{(\tilde{R}(\tilde\tau)^{3/n}+\tilde{\beta})^n}\nonumber \\
\tilde p(\tilde\tau)=\frac{3}{8\pi}\frac{-\tilde{\beta}}{(\tilde{R}(\tilde\tau)^{3/n}+\tilde{\beta})^{n+1}}
\end{align}
which shows that the stellar matter exhibits negative pressure. 
%This means that the strong energy condition may be violated from a certain radius of collapsing star.
It can be easily shown that the DEC
 holds however when $\tilde{R}<( 2 \tilde \beta)^{n/3}$ the SEC would be violated. This inequality is also reduced to (\ref{SEC}) and (\ref{SECB}) for Hayward and Bardeen BHs by choosing $n=1$ and $n=3/2$ respectively.
%\textbf{Moreover, the null, weak and dominant energy conditions are satisfied everywhere.} \\
The star surface is located at $\tilde{R}(\tilde{\tau})$ which can be obtained from (\ref{6dot}) and (\ref{alph}) as:
\[
 \frac{3}{2}\tilde \tau = \tilde R_0^{3/2} {}_2 F_1 [ -\frac{n}{2} , -\frac{n}{2}, 1-\frac{n}{2} ,- \tilde \beta  \tilde R_0^{-3/n}]-
\]
\begin{align}
\label{5last} 
 \tilde R^{3/2} {}_2 F_1 [ -\frac{n}{2} , -\frac{n}{2}, 1-\frac{n}{2} ,-\tilde \beta \tilde R^{-3/n}]
\end{align}
Setting $n=1$ and $n=3/2$, this agrees with previous results obtained in (\ref{7}) for Hayward and (\ref{0-9}) for Bardeen BHs. We have  plotted (\ref{5last}) in figure  \ref{nplot1}. At very small $\tilde\tau$, the evolution of the surface is independent of $n$ and $\tilde{\beta}$ and converges to the OS plot.
As the star’s radius decreases, we find that increasing $\tilde \beta$ slows down the rate of collapse for any value of $n$. 
Following, we will return to figure \ref{nplot1}, when dealing with horizons. 
%(However, fixing $\beta$, the rate of collapse
%initially decreases with increasing n... 
%In contrast, at late times decreasing $n$ slows down the rate of collapse, see figure \ref{nplot2}.\\
%In contrast, at late times the gravitational collapse speeds up with increasing $n$....)
%Analysis shows that for positive m the gravitational collapse is expected
%to slow down and if there is expansion, then it speeds up.
% By fixing $n$, we find that increasing $\tilde\beta$ slows down the rate of collapse whereas it is sensitive to the variation of $n$ only at the late stage of %gravitational collapse by fixing $\tilde\beta$. 
\begin{figure}
\centering
\includegraphics[width=3in]{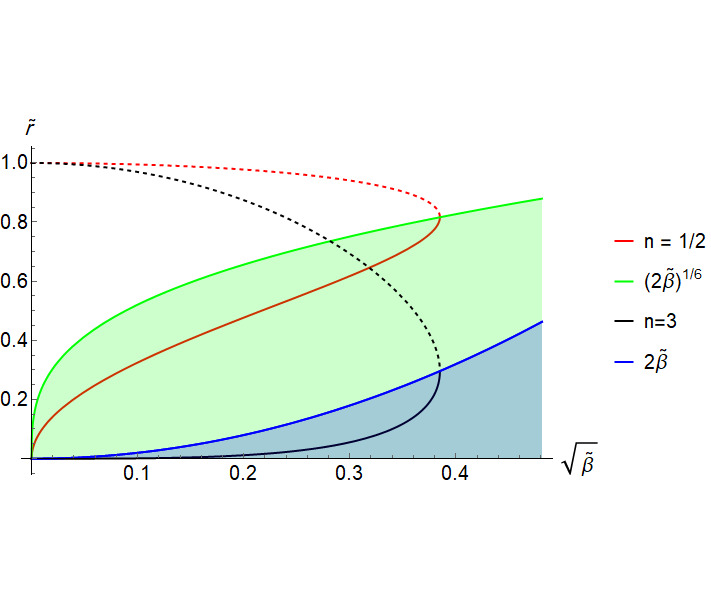}
\caption{$\tilde r_+$ and $\tilde{r_-}$ as functions of $\tilde\beta$. Dashed and solid curves represent $\tilde r_+$ and $\tilde r_-$ respectively.
 In the shaded region the SEC is violated.}
\label{allrp}
\end{figure}
The line element (\ref{5last}) has two event horizons located at
\begin{align}
\tilde{r}_{\pm}=\left(\frac{1}{3} +\frac{2}{3} \cos(\frac{\pi}{3} \mp \frac{1}{3}\cos^{-1} (\frac{27\tilde{\beta}}{2}-1))\right)^n
\end{align}
%\tilde{r}_{\pm} &=  \bigg[\frac{1}{3} + \frac{2}{3} \cos ( \frac{1}{3} \cos^{-1} ( 1 - \frac{ 27 \tilde\beta^2}{2}) + \frac{ 2 m_{\pm} \pi}{3})
%in which $m_{+}=0$ and $m_{-}=2$. 
if $\tilde\beta<4/27$. These are plotted in figure  \ref{allrp}. Also, it is shown that for a given $\tilde\beta$, the radius where the SEC is violated, decreases with increasing the polytropic index. \\
%\textbf{We may also note that we face the violation of SEC because of the negative pressure which is the reason that the BH does not collapse to a singularity. %Besides, the parameter $\tilde \beta$ determines the region. In other words, choosing the larger $\tilde \beta$, leads to smaller area where SEC is violated.}\\
As before, the interior apparent horizon can be extracted from (\ref{6dot}) and (\ref{alph}). These relations yield $R^{3/2}=((-\tilde{H})^{-2/n}-\tilde\beta)^{n/2}$ which can be substituted in (\ref{5last}) leading to the time evolution of apparent horizon as
\begin{align}\label{final}
 \frac{3}{2}\tilde \tau &=  \left((-\tilde H_0)^{-2/n}- \tilde\beta\right)^{n/2} \times \nonumber\\
&{}_2 F_1 [ -\frac{n}{2} , -\frac{n}{2}, 1-\frac{n}{2} ,-\frac{\tilde\beta}{ (-\tilde H_0)^{-2/n} - \tilde\beta}]\nonumber\\
& -   (\tilde r_{ah}^{2/n} - \tilde\beta)^{n/2} {}_2 F_1 [ -\frac{n}{2} , -\frac{n}{2}, 1-\frac{n}{2} , -\frac{\tilde\beta}{ \tilde r_{ah}^{2/n} - \tilde\beta}]
\end{align}
It has the following normal vector
\begin{equation}
n_{\alpha}=\left(\frac{3}{2}, \frac{{\tilde r_{ah}}^{2/n}}{{\tilde r_{ah}}^{2/n}-\tilde\beta}\right) 
\end{equation}
%As mentioned before the radius of the apparent horizon smoothly reaches a finite non-zero value. Depend on the polytropic index, the value of %reduction can change. For smaller values of polytropic index, the radius tends to a finite values smoother (see figure \ref{nah}).
with
\begin{align}
n_\alpha n^\alpha & \varpropto \frac{\tilde r_{ah}^{2/n}+3\tilde\beta}{\tilde r_{ah}^{2/n}-\tilde\beta}
\end{align}
which remains positive during the collapse. This comes from the fact that according to  (\ref{6dot}) and (\ref{alph}) $\tilde r_{ah}^{2/n}=\tilde R^{3/n}+\tilde\beta$ and hence $\tilde r_{ah}^{2/n}>\tilde\beta$.

Back to the outgoing null geodesics of (\ref{9}) mentioned before, it follows from (\ref{6dot}) and (\ref{alph}) that the event horizon satisfying 
\begin{equation}
\frac{d \tilde r_{eh}}{d \tilde R}-\frac{\tilde r_{eh}}{\tilde R}+\sqrt{\tilde R(1+\tilde \beta\tilde R^{-3/n})^n}=0
\end{equation}
Therefore
\[
\tilde{r}_{eh}=\tilde{R} + \tilde\beta^{n/2} \left({}_2 F_1 [-\frac{n}{2} ,- \frac{n}{3} ,1-\frac{n}{3}, -\frac{\tilde R^{3/n}}{\tilde\beta}]-\right.
\]
\begin{align}\label{finall}
\left.\frac{\tilde{R}}{\tilde{r}_+} {}_2 F_1 [-\frac{n}{2} ,- \frac{n}{3} ,1-\frac{n}{3}, -\frac{\tilde r_+^{3/n}}{\tilde\beta }]\right)
\end{align}
In figure  \ref{nplot1}, we have plotted the interior apparent and event horizons by setting $n=5/3$ for $\tilde \beta=0.01, 0.09$ and setting $n=5/2$ for $\tilde \beta=0.01, 0.09$. We see that for a fixed value of $n$,  the apparent (event) horizon decays (grows) more rapidly with decreasing $\tilde \beta$. 
% As it can be seen, the asymptotic values of both horizons decrease with increasing $n$. 
%Using (\ref{final})and (\ref{finall}), one can show that both apparent and event horizons go to $\tilde\beta^{n/2}$ at late times.
\section{Concluding Remarks}\label{Concluding Remarks}
In this paper, we studied some regular BHs with asymptotically deSitter core, formed by gravitational collapsing of a star. We assumed that the star collapses in free fall from infinity. Moreover, its interior geometry is described by spatially flat FRW spacetime which can be matched smoothly with an exterior regular BH, i.e. there is no energy shell separating them.
% This model deals with collapsing a uniform density sphere in the presence of anisotropic pressures to a regular BH. 
This is a generalization of OS collapse.\\
We first constructed a regular BH by inserting an arbitrary function in the Schwarzschild metric to regularize its singularity and discussed some general properties of star collapsing in this space--time.  Then we studied in details the generalized OS collapse to the two special regular BHs, Hayward and Bardeen. This includes the evolution of the star radius, its apparent and event horizon, and also the energy conditions, equation of state, and mass relation of the star. Interestingly, both mentioned BHs resulted from collapsing a polytropic star, and moreover the SEC is violated when the star radius becomes smaller than a specific value.\\
It is known that the polytropic equation of state plays an important role in the stellar structure models. This motivated us to find the general regular BH geometry  resulted from the collapse of a polytropic star. This yields to the line element (\ref{end0}) which reduces to the Hayward and Bardeen BHs for special values of polytropic index.\\
To get a physical intuition of regular BH (\ref{end0}), we investigated the photon spheres and found that there are two photon spheres where the smaller one is stable.\\
Here, a comparision can be made between the standard gravitational OS  collapse with the generalized one presented in this work:
\begin{itemize}
\item{In the generalized OS collapse, unlike the standard one, the collapsing process is ended in an infinite time, therefore the singularity is avoided.}
\item{The stellar matter should be have uniform density and pressure to get a smooth transition from a collapsing FRW interior geometry to a regular exterior  BH. This is unlike the standard OS collapse for which the stellar matter is pressureless.}
\item{In the generalized OS collapse, when the star radius becomes smaller than a specific value, the stellar matter violates the SEC. This radius is located somewhere between the inner and outer horizons of BH.}
\item{Unlike the OS collapse, in the generalized one, the apparent horizon radius does not go to zero necessarily and when the star contracts sufficiently, the trapped surfaces may be disappeared.}
\item{The event horizon has a similar evolution in the OS collapsing model and the generalized one. However in the generalized one, the rate of growing of event horizon depends on the free parameters of regular BH.}
\end{itemize} 
Although we have dealt here with regular BHs formed through gravitational collapse of a polytropic star, similar calculations can be carried out for other types of regular BHs. As an example, consider Kiselev metric \cite{kiselev}
\[
ds^2=-\left[1-\frac{2m}{r}-\frac{c}{r^{3\omega+1}}\right]dt^2+
\]
\begin{equation}\label{unreg}
{\left[1-\frac{2m}{r}-\frac{c}{r^{3\omega
+1}}\right]}^{-1}dr^2+r^2d\Omega^2
\end{equation}
This is a BH solution of Einstein equations in the presence of an anisotropic perfect fluid characterized by arbitrary parameters $c$ and $\omega$. Here, we propose the following regular Kiselev BH
\[
ds^2=-\left[1-\frac{2mr^2}{r^3+4ml^2}-\frac{c r^2}{r^{3(\omega
+1)}+2cl^2}\right]dt^2+
\]
\begin{equation}\label{regk}
{\left[1-\frac{2mr^2}{r^3+4ml^2}-\frac{cr^2}{r^{3(\omega
+1)}+2cl^2}\right]}^{-1}dr^2+r^2d\Omega^2
\end{equation}
in which we have regularized each term in metric components (\ref{unreg}) by adding a length scale parameter $l$.  As expected, for small (large) enough radial distance, the above metric approaches deSitter (Kiselev) space--time. It is notable that for $\omega=1/3$, metric (\ref{regk}) gives a regular Reissner-Nordstrom BH \cite{riazi}.\\
Performing the procedure followed in this paper, one can see that the collapsing star which produces the space--time (\ref{regk}), contains two components with different equation of state. Let us only mention here the result
\[
\tilde{\rho}_1=\frac{3}{8\pi}\frac{1}{\tilde{r}^3+2\tilde{l}^2} \hspace{1cm} {\tilde{\rho}}_2=\frac{3}{8\pi}\frac{\tilde{c}}{\tilde{r}^{3(1+\omega)}+2\tilde{c}\tilde{l}^2}
\]
\[
\tilde{p}_1=-\frac{16\pi}{3}\tilde{l}^2\tilde{\rho}_1^2
\]
\begin{equation}
\tilde{p}_2=-\frac{8\pi}{9c}\tilde{\rho}_2^2\left(6c\tilde{l}^2-3\omega(\frac{3\tilde{c}}{8\pi\tilde{\rho}_2}-2\tilde{c}\tilde{l}^2)^{\frac{2+3\omega}{3(1+\omega)}}\right)
\end{equation}
The first component is a polytropic matter and the second has a complicated equation of state. Each component corresponds to a term in metric (\ref{regk}). It should be noted that it is possible to use a single component stellar fluid however the analysis can not be done analytically. \\
Further investigation of our collapsing model and its application to the other geometries will be presented in the forthcoming paper.\\

{\bf Acknowledgments:}

The authors would like to thank the anonymous Referees for their careful reading of the paper and insightful comments and suggestions.
F. Shojai is grateful to the University of Tehran for supporting this work under a grant provided by the university research council.

%-------------------------

\end{document}